\documentclass[12pt,preprint]{aastex}

\newcommand{\asta}{A\&A}

\shorttitle{BeppoSAX Average spectra of Seyfert galaxies}
\shortauthors{Malizia et al.}

\begin{document}


\title{BeppoSAX average spectra of Seyfert galaxies}


\author{A. Malizia\altaffilmark{1}, L. Bassani\altaffilmark{1},
J. B. Stephen\altaffilmark{1} and G. Di Cocco\altaffilmark{1}}
\affil{IASF, Via Gobetti 101, 40129 Bologna, Italy}
\email{angela.malizia@bo.iasf.cnr.it}

\author{F. Fiore\altaffilmark{2}}
\affil{Osservatorio Astronomico di Roma, Via dell'Osservatorio,
I--00044 Monteporzio Catone, Italy }

\and

\author{A.J.Dean \altaffilmark{3}}
\affil{Physics Department, University of Southampton, Highfield, Southampton S09-5NH, UK}

\begin{abstract}
We have studied the  average 3-200 keV spectra of Seyfert  galaxies of  type 1 and 2, 
using  data  obtained with  BeppoSAX.  The average  Seyfert 1 spectrum is well-fitted by 
a power law continuum with photon  spectral index $\Gamma\sim$ 1.9, a 
Compton reflection component R$\sim$0.6-1 (depending on the inclination angle between 
the line of sight and the reflecting material)  and  a high-energy cutoff  at around 200 keV; 
there is also an iron line at 6.4 keV characterized by an equivalent width of  120 eV.   
Seyfert 2's on the other hand  show stronger neutral absorption (N$_{H}$=3-4 $\times$ 10$^{22}$ 
atoms cm$^{-2}$)  as expected but are also characterized by an  X-ray power law which 
is substantially harder  ($\Gamma$ $\sim$ 1.75) and with a cut-off at lower energies 
(E$_{c}\sim$130 keV);
the iron line parameters are instead substantially similar to those measured in 
type 1 objects.
There are only two possible solutions to this problem:
to assume more reflection in Seyfert 2 galaxies than observed in Seyfert 1
or more complex absorption than estimated in the first instance. The first
possibility is ruled out by the Seyfert 2 to Seyfert 1 ratio while the second 
provides an average Seyfert 2 intrinsic spectrum very similar to that of the Seyfert 1.
The extra absorber is likely an artifact due to summing spectra with different amounts
of absorption, althought we cannot exclude its presence in at least some  
individual sources.
Our result argues strongly for a very similar central engine  in both
type of galaxies as expected under the unified theory.

\end{abstract}


\section{Introduction}
The discovery of broad emission lines in the polarized spectrum of a
number of Seyfert (Sey) 2 galaxies has provided the basis for a
unified model of Sey galaxies in which the main discriminating
parameter is the inclination of our line of sight with respect to an
obscuring torus surrounding the source.  In the simple version of this
model, one then expects the direct continuum of type 1 and 2 objects to
be identical except for the effects of obscuration by the torus
material. The X-ray spectra of type 1 objects are relatively well
known and consist of a power law of index $\sim$2, a cut-off at around
200 keV plus a Compton reflection component with R (the solid angle in
units of 2$\pi$, subtended by the reflecting material) $\sim$ 0.6-1
and a Fe K$\alpha$ feature (Perola et al. 2002, Matt 2001, Pounds and Reeves
2002).  Conversely, the X-ray spectra of type 2 are much
less well defined due to the extra complexity of absorption:
constraints on individual spectra are rather poor due to the limited
photon statistics and provide a greater variety of spectral components
(Matt et al. 2000, Malaguti et al. 1999, Turner et al. 2000). One
way to better constrain the X-ray spectral properties is to consider
average spectra.  When this is done, the spectra of type 2 objects
are found to be substantially harder than that of type 1, which implies a
lower energy cut-off (Zdziarski et al. 1995, 2000).
However, this intrinsic difference in the spectrum of Sey 1 and 2's
cannot be explained by the simplest version of the unified AGN model,
providing its strongest challenge yet. 
Here, we study the average spectra
of Sey 1 and 2 observed by BeppoSAX over the energy band 3-200
keV, i.e.  the first simultaneous broad band spectra
with coverage above 10 keV.
The purpose of this work is to impose the tightest possible constraints on
both spectral shapes in order to check if their intrinsic continuum is indeed
different and to discuss our findings in the framework of the unified
theory.

\section{The average spectra}

The sample used for this work consists of 9 Sey 1 (Fairall9, 3C120,
NGC3783, NGC4593, MCG-6-30-15, NGC5548, MKN509, ESO141-G55,
NGC7469) and 13 Sey 2 (NGC526A, NGC2992, NGC5506, ESO103-G35, NGC7172,
NGC7314, NGC7582, MCG-5-23-16, NGC1365, NGC2110, NGC4258, NGC5252,
NGC5674) galaxies. The strongest BeppoSAX detections of Sey (such as
NGC4151 and IC4329A) were excluded so that single sources would not
significantly bias our results. We also excluded from our Sey 2
samples sources with column density in excess of 10$^{23}$ atoms
cm$^{-2}$ in order to maintain a certain homogeneity in the absorption
properties.  The count spectra of the individual observations were added with
weights corresponding to the length of time of each observation and the
sum was simply normalized by the total number of sources in each subset. 
Only MECS (3-10 keV) and PDS data (15-200 keV) were considered in order to
avoid the complexity due to the soft excess particularly in Sey 2.
Standard data reduction was performed using the software package
"SAXDAS" (see http://www.sdc.asi.it/software and the Cookbook for
BeppoSAX NFI spectral analysis, Fiore, Guainazzi \& Grandi 1998).
Data were linearized and cleaned for Earth occultation periods and
periods of high particle background (satellite passages
through the South Atlantic Anomaly).  Data have been accumulated for
Earth elevation angles $>5$ degrees and magnetic cut-off rigidity
$>6$. For the PDS data we adopted a fine energy and temperature
dependent Rise Time selection, which decreases the PDS background by
$\sim 40 \%$. This improves the signal to noise ratio of faint sources
by about 1.5 (Frontera et al. 1997, Fiore,
Guainazzi \& Grandi 1998).  Data from the four PDS units and the two
MECS units were merged after equalization and single MECS and PDS
spectra were then accumulated.  The normalization constant, introduced
to allow for a well known difference in the absolute cross-calibration
between the two detectors, was left free to vary between 0.7 -0.95
(Fiore, Guainazzi and Grandi, 1998). 
All quoted errors refer to 90\% confidence level for 1 interesting parameter 
($\Delta \chi^{2}$=2.7).
First we fitted the 3-200 keV spectra with an absorbed power law plus a
gaussian line (model 1 in table 1). As expected, this simple model
provides an unsatisfactory fit to the data mainly due to the presence
of a strong reflection component and the possible presence of a high energy cut-off.
The iron line energy is compatible with reflection in cold material and the
equivalent width (EW) is $\sim$ 100-200 eV. In both classes
of objects the line width is slightly broad but this is likely due to  a redshift
effect as sources with different z are summed together: in particular
5 Seyfert 1s and 3 Seyfert 2s galaxies have redshift greater than 0.01
which is enough to produce a broadening of the line; an important
contribution to the line width could also be given by Fairall 9 (z=0.047) to the Sey 1
average spectrum and by NGC5506 (z=0.062) to the Seyfert 2s average spectrum.
A detailed analysis of the iron line is however beyond the scope of the 
present paper and it is therefore  postponed to a later work.\\
Although the absorbed power law  model points to an overall shape similar 
for both types
of galaxies, the best fit photon index for Sey 1 and 2 turns out to be
significantly different: the Sey 1s retain
the canonical $\Gamma$ of $\sim$1.9 while the Sey 2s are flatter with $\Gamma \sim$
1.75.  As expected Sey 2 galaxies are more absorbed than Sey 1 with an
average column density of $\sim$ 4 $\times$ 10$^{22}$ atoms cm$^{-2}$,
however some amount of absorption is also present in Sey 1 galaxies
with N$_{H}$ $\simeq$ 10$^{22}$ atoms cm$^{-2}$. A recent analysis of a
sample of Sey 1 with BeppoSAX indicates that cold absorption
of this amount is generally not present in this type of object 
(N$_{H}$ $\le$ 10$^{21}$ atoms cm$^{-2}$); however a warm absorber 
seems to be a common feature of type 1 objects and is consistent with the presence
of some absorption in our average spectrum (Perola et al. 2002). Next 
we introduced a
Compton reflection component  and a high energy cut-off via the pexrav model in
XSPEC: this model provides an improvement in the fit which is
significant at the 99.99\% confidence level for 2 extra degrees of freedom
for both classes (table 1 model 2).  The inclination assumed for the
Seyfert 1s is cos i=0.45 in model 2$^{a}$ and 0.87 in model 2$^{b}$; 
on the other hand since the Sey 2s
are most likely to be seen edge-on, we assumed in this case cos i=0.45 only.
Furthermore, we assume the reflection medium to be close to neutral and the
abundances used are those of Anders and Ebihara (1982).  
It is evident from table 1 that while the amount of
reflection is consistent between the two types, the primary continuum
is still significantly different: the average Sey2 spectrum is
substantially harder than that of Sey 1 and the cut-off is at lower
energies (see table 1).  On the other hand, the average spectrum of Sey 1s is
perfectly canonical having  $\Gamma$=1.9, a reflection strength R $\sim$ 0.6 (or 1
depending on the inclination angle assumed) and a cut-off energy at around 200 keV;
the iron line is confirmed at 6.4 keV and the equivalent width is 100-120
keV. Our results confirm the findings of Perola et al. (2002) on a sample of  
Seyfert 1 galaxies observed by BeppoSAX and analyzed individually with the same model.
When their data are combined to get weighted mean values ($\Gamma$=1.84$\pm$0.02, R=0.64$\pm$0.09 
for cos$\Theta$=0.9 and E$_{\rm cut}$ 130$\pm$20 keV), these are very similar to our 
best fit parameters of model 2$^{b}$ in table 1;
their average line energy is at 6.44$\pm$0.04 keV while the equivalent width is 122$\pm$16
again in agreement with our results.
Also, if the iron line
comes from the same medium as the Compton reflecting continuum, one would expect
a simple relationship between R, $\Gamma$ and EW, which for the average photon index
of the Sey 1 and the reflection value obtained for cos$\Theta$=0.87 
provides an EW value very similar to that observed (see George and Fabian 1991
results rescaled to the Anders and Grevesse (1993) abundances): therefore the observed 
EW is of the magnitude expected from the simultaneously measured strength
of the reflection component.
The difference seen between
Sey 1 and 2 is however troublesome as it is not easily reconciled with the 
unified theory, which is however independently supported by other observational
evidences.\\  
It is well known that there is an inter-dependence of the parameters in the pexrav
model, whereby increasing the reflection and/or the high energy cut-off leads to
a steeper power law.
To test this possibility, we have imposed a photon index of 1.9 (equal
to that obtained for the Sey 1 in the cos i=0.45 configuration) on the Sey
2 and measured the contours of reflection strength against cut-off
energy for this case: figure 1 shows the results compared to
the Sey 1 case while model 2$^{a,c}$ in table 1 provides the best fit
parameters.  The cut-off energy turns out to be similar
to that measured in type 1 objects but R is now much higher ($\sim$3). 
A Compton reflection component as strong as this is unusual for Compton thin 
Sey 2 galaxies (Bassani et al. 1999) which characterize our sample.
Also, following the same reasoning as for type 1 objects, the measured iron line EW
is too low for the observed reflection strength especially  if a
contribution of the order of 30-40  eV has to be taken into  account for the line
component originating by transmission through a torus of column density of 
$\sim$ 4 $\times$ 10$^{22}$ cm$^{-2}$.
A  reflection component as strong as that
measured is,  however, still possible if the hard X-ray emission is isotropic 
and/or there is a time lag in the response of the reflection flux to variations 
of the source intensity (Cappi et al. 1996, Weaver et al. 1996). \\ 
A model independent way to verify the presence of a stronger reflection in
type 2 with respect to type 1 objects 
is to measure the Sey 2 to Sey 1 PDS ratio and search for the presence of
a bump in the PDS data.
The Sey2 to Sey1 ratio is shown in figure 2, which clearly excludes the presence 
of stronger reflection in type 2 objects. Indeed the average spectra of Sey 1s and Sey 2s
appear to be very similar in the high energy (PDS) band, but they are different 
in the MECS range where effect of absorption takes place.
\\
The ratio shown in figure 2 was produced by first summing the background-subtracted 
data for the two types of sources individually, normalizing each for the number of
sources in each sample and then dividing in order to obtain the ratio.
The data were then rebinned in order to have a minimum (3 sigma) detection in
each energy bin throughout the ratio.
A visual inspection of the distribution of spectra for each sample (figures 3)
confirms that neither the Sey 1 or Sey 2 groups are dominated by one source.
The origin of the negative counts which can be seen in the individual spectra is due to 
the fact that we are working in a high energy regime where we are background dominated 
and statistical variations in background can be higher than the source flux itself.
This is confirmed by the observation that the maximum negative flux corresponds to
a significance of only 1.5$\sigma$, thus ruling out the possibility that systematics in the
background subtraction are responsible.
The Sey 1 sample is much more homogeneous than the Sey 2 group, which would be
expected due to the wider range of geometric orientations possible for the latter type.
It can be seen that there is no discernible excess in the 20-100 keV band, 
thus excluding the
presence of stronger reflection in Sey 2's than in Sey 1's; if we restrict the analysis 
to the 40-70 keV band where excess counts are visible by eye, the significance is still
too low (2.3$\sigma$) to claim the presence of an excess.\\
Having excluded reflection as the possible cause of a flatter
spectrum, the only other remaining way to steepen the Sey 2 spectrum 
is via a more complex absorption than assumed.
This is not an unreasonable assumption since within our sample there are
objects with different column densities and summing their spectra together can 
artificially produce a complex absorption where the covering fraction depends on the 
distribution of fluxes.
We have therefore introduced in the Sey 2 fit an extra absorber that partially
covers the source (model 3$^{a}$ in Table 1). 
The fit improves
significantly at more than 99\% confidence level for two extra degrees of freedom
and provides average spectral parameters for the intrinsic continuum 
very similar to those measured in Sey 1s: $\Gamma \sim$  1.9, R $\sim$0.7-1.2 and E$_{c}\sim$ 200 keV.
The two best fits give rise to two different MECS/PDS normalization factors ($\sim$0.75 and $\sim$0.83)
both of which are within the range indicated in the Cookbook of BeppoSAX NFI spectral analysis
(Fiore et al. 1998). However, in order to further check the robustness of these results, we
have tried to re-analyse the Sey 2 data using the normalization found for the Sey 1's and vice versa.
In both cases the values of $\Gamma$, R and cut-off are consistent with those found
in the original analysis within 2$\sigma$. 
The composite spectra of
both types of Seyferts are compared in figure 4; again from this figure is evident that 
the two spectra are very similar in the PDS energy range, while differing considerable at 
lower energies.

\section{Conclusions}

We have measured with BeppoSAX the average spectra of Sey 1 galaxies
confirming previous indications that typically these galaxies have a
power law shape with photon index $\Gamma \sim$ 1.9, a reflection component of
strength R$\sim$0.6-1 and a high energy cut-off which is now more
clearly defined and located at around 200 keV. 
Sey 2s on the other hand,
have an average spectrum which is substantially harder and with a
lower energy cut-off than Sey 1s.
Our analysis excludes the possibility that stronger reflection in Sey 2s 
is responsible for their harder spectrum; instead the introduction of a 
complex absorber provides an average Sey 2's spectrum very similar to that
of Sey 1's. This complex absorption is not necessarily intrinsic to individual
sources but is likely to be an artifact of summing spectra with different
amounts of absorption. However since complex absorption has been observed in
some individual source spectra (e.g. Malaguti et al. 1999, Turner et al. 2000)
we cannot exclude a-priori its possible contribution to the average Seyfert 2's
spectrum.  
In any case our result  on the primary power law continuum argues strongly 
for a very similar production
mechanism not only in both types of Seyferts but in each individual
object and is therefore also relevant for the synthesis of the X-ray
background: Sey 2 spectra can now be taken to be similar to their type
1 counterparts in the AGN contribution estimates.

\acknowledgments

We are grateful to M. Cappi for useful discussions. A.M. acknowledge financial
support from ASI.

\begin{figure}
\plotone{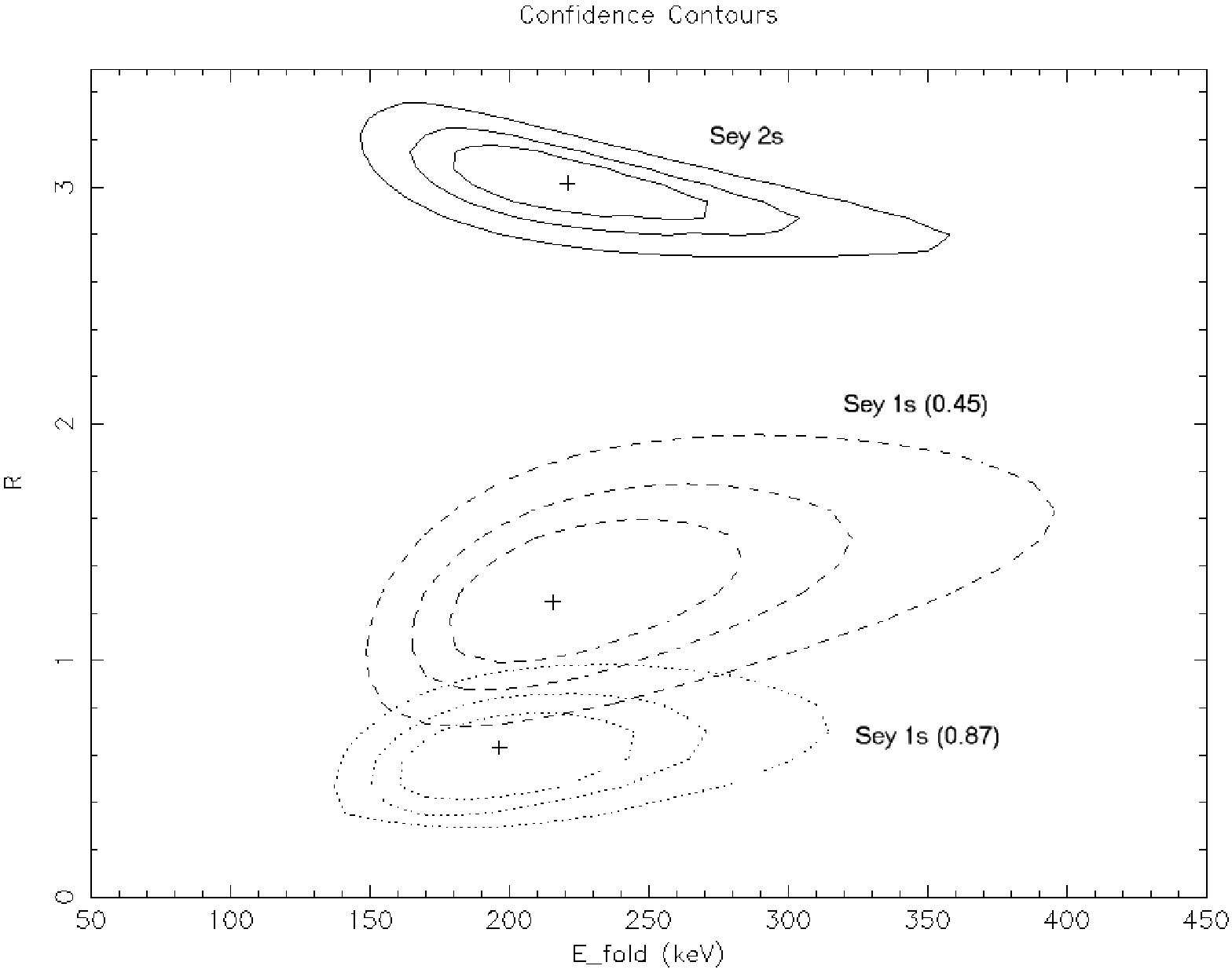}
\caption{Contour plot of reflection versus cut-off with the photon index fixed to 1.9 in the Sey 2 fit. 
\label{fig1}}
\end{figure}

\begin{figure}
\plotone{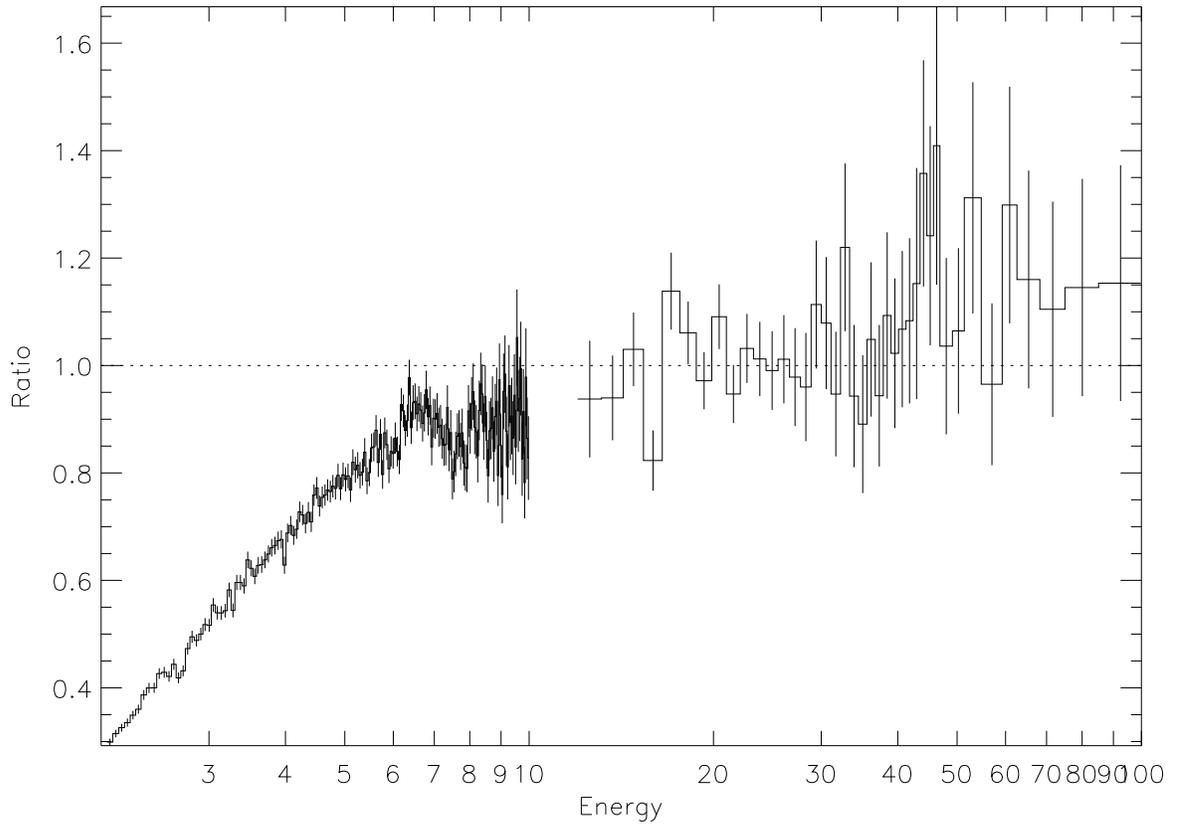}
\caption{The MECS-PDS ratio (Sey2/Sey 1) of the average normalized spectra. \label{fig2}}
\end{figure}

\begin{figure}
\plotone{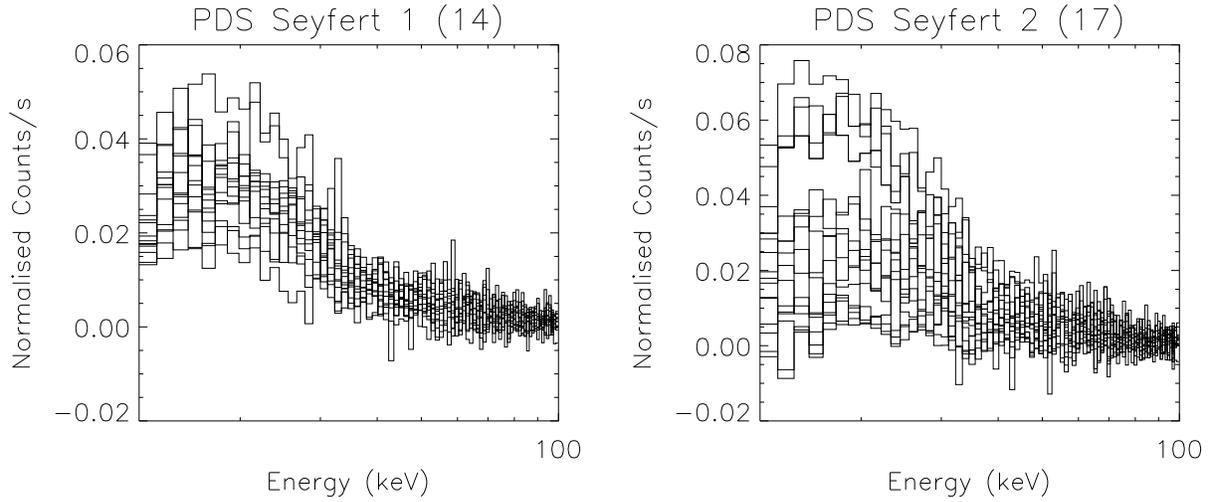}
\caption{The normalized spectra of Sey 1 sample in the PDS detector (left). The normalized 
spectra of Sey 2 sample in the PDS detector (right).
The negative counts present in each set of data are due to the fact that we are working in a high energy 
regime where we are background dominated and statistical variations in background can be higher
than the source flux itself.\label{fig3}}
\end{figure}

\begin{figure}
\plotone{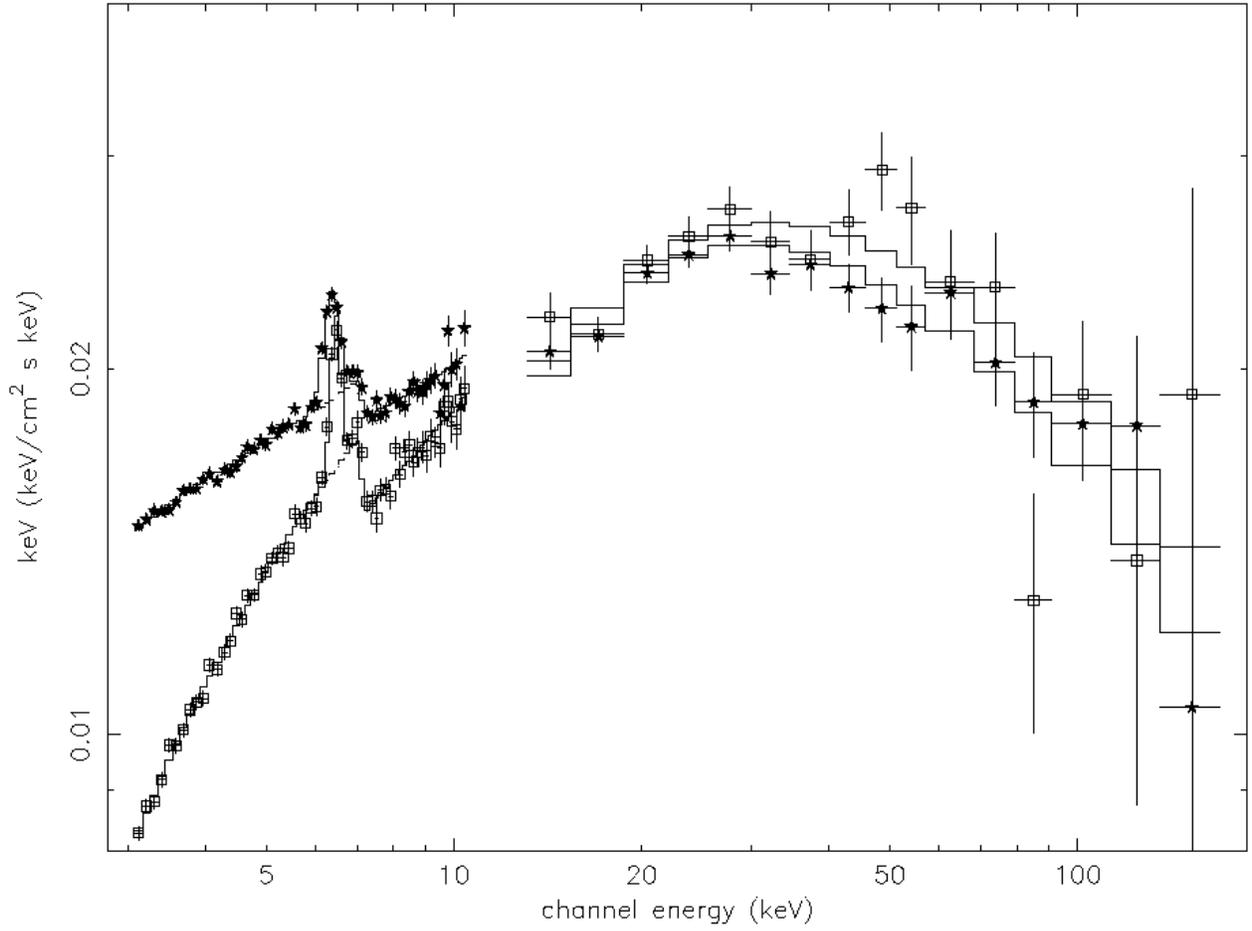}
\caption{Comparison of the unfolded spectra of Sey 1s (stars) fitted with pexrav model
and Sey 2s (squares) fitted with double absorption plus pexrav model \label{fig4}}
\end{figure}


\begin{deluxetable}{llllcllll}
\tabletypesize{\scriptsize}
\tablecaption{ \label{tbl-1}}
\tablewidth{0pt}
\tablehead{
\colhead{} &  \multicolumn{3}{c}{\bf Seyfert 1s}   & \colhead{} &  \multicolumn{4}{c}{\bf Seyfert 2s}   \\
\cline{2-4} \cline{6-9} \\
\colhead{Parameter} & \colhead{Model 1}   & \colhead{Model 2\tablenotemark{a}}   &
\colhead{Model 2\tablenotemark{b}} & \colhead{} &
\colhead{Model 1}  & \colhead{Model 2\tablenotemark{a}} & \colhead{Model 2\tablenotemark{a,c}} & \colhead{Model 3\tablenotemark{a}}} 
\startdata
N$_{H}$(1) $\times$ 10$^{22}$ & 1.01$^{+0.15}_{-0.17}$ & 0.93$^{+0.18}_{-0.18}$ & 0.87$^{+0.18}_{-0.19}$ &  
                              & 4.12$^{+0.23}_{-0.22}$ & 3.67$^{+0.29}_{-0.27}$ & 3.83$^{+0.24}_{-0.30}$
                              & 3.16$^{+0.50}_{-0.13}$ \\
N$_{H}$(2) $\times$ 10$^{22}$ & \nodata                & \nodata                & \nodata                &
                              & \nodata                & \nodata                & \nodata                
                              & 27$^{+5}_{-4}$   \\
C$_{f}$                       & \nodata                & \nodata                & \nodata                &
                              & \nodata                & \nodata                & \nodata
                              & 0.26$^{+0.04}_{-0.03}$   \\
Photon Index                  & 1.85$^{+0.01}_{-0.02}$ & 1.92$^{+0.04}_{-0.03}$ & 1.88$^{+0.04}_{-0.03}$ &
                              & 1.74$^{+0.02}_{-0.02}$ & 1.75$^{+0.06}_{-0.04}$ & {\bf 1.90}
                              & 1.89$^{+0.02}_{-0.13}$   \\
E$_{fold}$                    & \nodata                & 216$^{+75}_{-41}$      & 197$^{+64}_{-40}$      &
                              & \nodata                & 128$^{+45}_{-24}$      & 221$^{+94}_{-61}$
                              & 226$^{+69}_{-48}$        \\
R                             & \nodata                & 1.25$^{+0.38}_{-0.29}$ & 0.64$^{+0.22}_{-0.19}$ &
                              & \nodata                & 1.19$^{+0.64}_{-0.35}$ & 3.02$^{+0.28}_{-0.27}$
                              & 0.96$^{+0.29}_{-0.24}$   \\
E$_{line}$ (keV)              & 6.39$^{+0.04}_{-0.04}$ & 6.37$^{+0.03}_{-0.03}$ & 6.38$^{+0.04}_{-0.04}$ &
                              & 6.48$^{+0.06}_{-0.10}$ & 6.47$^{+0.03}_{-0.04}$ & 6.44$^{+0.04}_{-0.03}$
                              & 6.46$^{+0.03}_{-0.03}$ \\
$\sigma_{line}$  (keV)        & 0.26$^{+0.06}_{-0.05}$ & 0.18$^{+0.03}_{-0.14}$ & 0.22$^{+0.06}_{-0.08}$ &
                              & 0.25$^{+0.07}_{-0.02}$ & 0.21$^{+0.06}_{-0.11}$ & 0.10$^{+0.06}_{-0.10}$
                              & 0.12$^{+0.07}_{-0.12}$ \\
EW (eV)                       & 148$^{+19}_{-18}$      & 106$^{+18}_{-29}$      & 120$^{+21}_{-20}$      &
                              & 189$^{+26}_{-25}$      & 145$^{+71}_{-36}$      & 97$^{+17}_{-14}$
                              & 103$^{+13}_{-19}$  \\
$\chi^{2}$/dof                & 257.41 (70)            & 87.12 (68)             & 90.48 (68)             &
                              & 242.62 (70)            & 101.92 (68)            & 108.54 (69)
                              & 90.4 (66)           \\
$\chi^{2}_{\nu}$              & 3.68                   & 1.25                   & 1.33                   &
                              & 3.47                   & 1.50                   & 1.57
                              & 1.37                  \\
\enddata

\tablenotetext{a}{cos$\Theta$=0.45 (fixed)},
\tablenotetext{b}{cos$\Theta$=0.87 (fixed)},
\tablenotetext{c}{$\Gamma$=1.90 (fixed)}.

\tablecomments{{\bf Model 1} = wa$\star$po+ga (absorbed power law + gaussian line);
{\bf Model 2} = wa$\star$(pexrav+ga) (absorbed exponentially cut off power law spectrum reflected 
from neutral material);
{\bf Model 3} = wa$\star$pcfabs$\star$(pexrav+ga) (as Model 2 plus a partial covering fraction 
absorption).}
\end{deluxetable}

\end{document}